\documentstyle[preprint,aps]{revtex}

\tightenlines

\begin{document}

\draft

\title{
Masses, Magnetic Moments, and Semileptonic Decays of\\
Spin 1/2 Baryons with Sea Contribution
}

\author{
V.~Gupta$^{\rm a}$,
P.~Ritto$^{\rm a}$, and
G.~S\'anchez-Col\'on$^{\rm a,b}$
}
\address{
$^{\rm a}$Departamento de F\'\i sica Aplicada\\
Centro de Investigaci\'on y de Estudios Avanzados del IPN. Unidad M\'erida\\
A.P. 73, Cordemex, M\'erida, Yucat\'an 97310. MEXICO\\
$^{\rm b}$Department of Physics, University of California\\
Riverside, CA 92521-0413, U.S.A.
}

\date{March 23, 1998}

\maketitle

\begin{abstract}
The spin 1/2 baryons are pictured as a composite system made out of a ``core" of three valence quarks (as in the simple quark model) surrounded by a ``sea" (of gluon and $q\bar q$ pairs) which is specified by its total quantum numbers. We assume the sea is a $SU(3)$ flavor octet with spin 0 or 1 but no color. This model, considered earlier, is used to obtain simultaneous fits for masses, magnetic moments and $G_A/G_V$ for semileptonic decays. These fits give predictions for nucleon spin distributions in reasonable agreement with experiment.
\end{abstract}


\section{Introduction}
\label{i}

The simple quark model (SQM) though qualitatively successful, fails to account for low energy properties of baryons quantitatively. Experimentaly~\cite{1}, it is found that quarks cannot  even account for the proton spin and thus it is necessary to go beyond SQM. Since quarks interact through strong color forces mediated by gluons, a physical hadron, in reality, consists of valence quarks surrounded by a ``sea" of gluons and quark-antiquark ($q\bar q$) pairs. The effect of the sea contribution to hadron structure has been considered by several authors~\cite{2,3,4,5,6,7,8,9}. 

In this paper, we study the static properties of the spin 1/2 baryons ($p$, $n$, $\Lambda$, ...) following Refs.~\cite{8} and~\cite{9} where the general sea is specified by its total flavor, spin and color quantum numbers. The baryons are pictured as a composite system made out of a baryon ``core" of the three valence quarks (as in SQM) and a flavor octet sea with spin 0 and 1 but no color. Earlier~\cite{8,9}, such a physical baryon wavefunction was applied to baryon magnetic moments and semileptonic decays and gave excellent fits. The purpose of this paper is to use this wavefunction to obtain a simultaneous fit to masses, magnetic moments and semileptonic decays. 

Sec.~\ref{ii} gives the wavefunction for the physical baryon in our model. Sec.~\ref{iii} presents a discussion of the mass operator used and the models for the ``core" baryon masses. It also gives briefly how the magnetic moments and semileptonic decays are calculated. Sec.~\ref{iv} gives the combined fits to masses, magnetic moments and semileptonic decays, while a prediction of these fits for nucleon spin distributions is discussed in Sec.~\ref{v}. Lastly, Sec.~\ref{vi} gives some concluding remarks.

\section{Spin 1/2 Baryon Wave Functions with Sea}
\label{ii}

The physical baryon octet states, denoted by $B(1/2\uparrow)$ are obtained by
combining the ``core" wavefunction $\tilde{B}({\bf 8},1/2)$ (the usual SQM spin 1/2 baryon octet wave function) with the sea wavefunction with specific properties given below.
We assume the sea is a color singlet but has flavor and spin properties which
when combined with those of the core baryons $\tilde{B}$ give the desired
properties of the physical baryon $B$.
Since both the physical and core baryon have $J^P=\frac{1}{2}^+$, this implies
that the sea has even parity and spin 0 or 1.
The spin 0 and 1 wavefunctions for the sea are denoted by $H_0$ and $H_1$,
respectively.
We also refer to a spin 0 (1) sea as a scalar (vector) sea.
For $SU(3)$ flavor we assume the sea has a $SU(3)$ singlet component and an
octet component described by wavefunctions $S({\bf 1})$ and $S({\bf 8})$,
respectively.
The color singlet sea in our model is thus described by the wavefunctions
$S({\bf 1})H_0$, $S({\bf 1})H_1$, $S({\bf 8})H_0$, and $S({\bf 8})H_1$.

The total flavor-spin wavefunction of a spin up ($\uparrow$) physical baryon
which consists of 3 valence quarks and a sea component (as discussed above)
can be written schematically as

\begin{eqnarray}
B(1/2\uparrow)
&=&
\tilde{B}({\bf 8},1/2\uparrow)H_{0}S({\bf 1})+
b_{{\bf 0}}\left[\tilde{B}({\bf 8},1/2)
\otimes H_{1}\right]^{\uparrow}S({\bf 1})
\nonumber\\
&&
+\sum_{N}a(N)\left[\tilde{B}({\bf 8},1/2\uparrow)H_{0}
\otimes S({\bf 8})\right]_{N}
\label{1}\\
&&
+\sum_{N}b(N)\left\{[\tilde{B}({\bf 8},1/2)\otimes H_{1}]^{\uparrow}
\otimes S({\bf 8})\right\}_{N}.
\nonumber
\end{eqnarray}

\noindent
The first term is the usual $q^{3}$-wavefunction of the SQM (with a trivial
sea) and the second term (coefficient $b_0$) comes from spin-1 (vector) sea
which combines with the spin 1/2 core baryon $\tilde{B}$ to a
spin 1/2$\uparrow$ state.
So that,

\begin{equation}
\left[\tilde{B}({\bf 8},1/2)\otimes H_{1}\right]^{\uparrow} =
\sqrt{\frac{2}{3}}\tilde{B}({\bf 8},1/2\downarrow)H_{1,1} -
\sqrt{\frac{1}{3}}\tilde{B}({\bf 8},1/2\uparrow)H_{1,0}.
\label{2}
\end{equation}

\noindent
In both these terms the sea is a flavor singlet.
The third (fourth) term in Eq.~(\ref{1}) contains a scalar (vector) sea
which transforms as a flavor octet.
The various $SU(3)$ flavor representations obtained from
$\tilde{B}({\bf 8})\otimes S({\bf 8})$ are labelled by
$N={\bf 1,8_{F},8_{D},10,\bar{10},27}$.
As it stands, Eq.~(\ref{1}) represents a spin 1/2$\uparrow$ baryon which is
not {\em a pure flavor octet} but has an admixture of other $SU(3)$
representations weighted by the unspecified constants $a(N)$ and $b(N)$.
It will be a flavor octet if $a(N)=b(N)=0$ for $N={\bf 1,10,\bar{10},27}$.
The color wavefunctions have not been indicated as the three valence quarks in
the core $\tilde{B}$ and the sea (by assumption) are in a color singlet state. The sea isospin multiplets contained in the $SU(3)$ flavor octet $S({\bf 8})$ are denoted as $(S_{\pi^+},S_{\pi^0},S_{\pi^-})$, $(S_{K^+},S_{K^0})$, $(S_{\bar{K}^0},S_{K^-})$, and $S_{\eta}$.
The familiar pseudoscalar mesons are used here as subscripts to label the isospin and hypercharge quantum numbers of the sea states. Details of the wavefunction have been given earlier~\cite{8,9}. However, for completeness the explicit physical baryon states in terms of the core and sea states are given in Tables~\ref{tabla1} and \ref{tabla2}. The normalization of a given baryon state (not indicated in Eq.~(\ref{1})) depends on the parameters which enter in the wavefunction and is different for different isospin multiplets (see Table~\ref{tabla3}).

For our applications we adopt the phenomenological wavefunction given in
Eq.~(\ref{1}), where the physical spin 1/2 baryons have admixtures of flavor
$SU(3)$ determined by the coefficients $a(N)$ and $b(N)$,
$N={\bf 1,10,\bar{10},27}$.
As we shall see, such a wavefunction which respects the isospin and
hypercharge properties of the usual spin 1/2 baryon states is general enough
to provide an excellent fit to the masses, magnetic moments and semileptonic decays data simultaneously. Only few of the thirteen parameters in Eq.~(\ref{1}) are needed for this purpose.

For applications, we need the quantities $(\Delta q)^B$, $q=u,d,s$; for each
spin-up baryon $B$. These are defined as

\begin{equation}
(\Delta q)^B =
n^B(q\uparrow)-n^B(q\downarrow)+n^B(\bar{q}\uparrow)-n^B(\bar{q}\downarrow),
\label{4}
\end{equation}

\noindent
where $n^B(q\uparrow)$ ($n^B(q\downarrow)$) are the number of spin-up
(spin-down) quarks of flavor $q$ in the spin-up baryon $B$.
Also, $n^B(\bar{q}\uparrow)$ and $n^B(\bar{q}\downarrow)$ have a similar
meaning for antiquarks.
However, these are zero as there are no explicit antiquarks in the
wavefunctions given by Eq.~(\ref{1}).
The expressions for $(\Delta q)^B$ reduce to the SQM values if there is no sea
contribution, that is, $b_0=0$, $a(N)=b(N)=0$,
$N={\bf 1,8_F,8_D,10,\bar{10},27}$.

\section{Magnetic Moments, Masses and Semileptonic Decays}
\label{iii}

For any operator $\hat{O}$ which depends only on quarks, the matrix elements
are easily obtained using the orthogonality of the sea components.
Clearly $\langle B\uparrow|\hat{O}|B'\uparrow\rangle$ will be a linear
combination of the matrix elements
$\langle \tilde{B}\uparrow|\hat{O}|\tilde{B'}\uparrow\rangle$ (known from
SQM) with coefficients which depend on the coefficients in the wavefunction, Eq.~(\ref{1}).

\subsection{Magnetic Moments (MM's)}
\label{iiia}

We assume the baryon magnetic moment operator, $\hat{\mu}$, to act solely on the valence quarks in $\tilde B$, so that

\begin{equation}
\hat{\mu}\equiv \sum_{q} \mu_q \sigma_z^q  
\label{5}
\end{equation}

\noindent
where $\mu_{q}=e_{q}/2m_{q}$ and $e_q$ and $m_q$ are quark charge and mass for $q=u,d,s$.

It is possible to show that the MM's of the spin 1/2 baryons, $\mu_{B}$ ($B=p,n,\Lambda,\dots$), and the transition magnetic moment, $\mu_{\Sigma^{0}\Lambda}$, can be written as

\begin{equation}
\mu_{B}=\sum_{q=u,d,s}(\Delta q)^{B}\mu_{q}
\qquad \hbox{and} \qquad
\mu_{\Sigma^{0}\Lambda}=\sum_{q=u,d}(\Delta q)^{\Sigma^{0}\Lambda}\mu_{q},
\label{6}
\end{equation}

\noindent
where the $(\Delta q)^{B}$ are defined in Eq.~(\ref{4}). Expressions for $(\Delta q)^{B}$ in terms of the parameters $b_0$, $\beta_i$ and $\beta_i^\prime$ are given in Ref.~\cite{8}.
From Eqs.~(\ref{6}) we see that the MM's depend on the quark masses
(or quark MM's) and on the parameters $b_0$, $a(N)$, $b(N)$ which determine the sea. 

\subsection{Masses}
\label{iiib}

For masses we assume that the mass operator $H$ acts only on the quarks in the core $\tilde B$, this gives the physical baryon masses

\begin{equation}
m_{B}=\sum_{\tilde B} \Omega_{B\tilde B}m_{\tilde B}
\label{8}
\end{equation}

\noindent
as a linear combination of the eight ``core" baryon masses $m_{\tilde B}$ weighted by the coefficients $\Omega_{B\tilde B}$ (given in Table~\ref{tabla3}) which depend on the parameters of the wavefunction, Eq.~(\ref{1}).

The parameters in the wavefunction can be fixed by fitting other data ({\em e.g.} MM's) and thus determine $\Omega_{B\tilde B}$. However, we still need to know $m_{\tilde B}$ to be able to calculate $m_B$. For this purpose, we assume the mass operator of the form

\begin{equation}  
{H}=H_0+H_8+H_3,
\label{9}
\end{equation}

\noindent
where $H_0$ is flavor $SU(3)$ singlet and $H_8$ transforms like the eighth component of an octet and breaks flavor $SU(3)$ down to $SU(2)_{I}\otimes U(1)_{Y}$. The last term $H_3$ transforms like $I=1$, $I_3=0$ or third component of an octet. It breaks $SU(2)_{I}$ giving different masses to members of an isospin multiplet in $\tilde{B}({\bf 8})$.

Given these general transformation properties for ${H}$, one can express the eight masses of the core baryon octet as

\begin{eqnarray}
m_{\tilde{p}}&\equiv& m_0-(F_8+D_8)-(F_3-D_3),
\nonumber\\
m_{\tilde{n}}&\equiv& m_0-(F_8+D_8)+(F_3-D_3),
\nonumber\\
m_{\tilde{\Lambda}}&\equiv& m_0-2D_8,
\nonumber\\
m_{\tilde{\Sigma}^{+}}&\equiv& m_0+2D_8-2F_3,
\nonumber\\
m_{\tilde{\Sigma}^{0}}&\equiv& m_0+2D_8,
\label{10}
\\
m_{\tilde{\Sigma}^{-}}&\equiv& m_0+2D_8+2F_3,
\nonumber\\
m_{\tilde{\Xi}^{0}}&\equiv& m_0+(F_8-D_8)-(F_3+D_3),
\nonumber\\
m_{\tilde{\Xi}^{-}}&\equiv& m_0+(F_8-D_8)+(F_3+D_3),
\nonumber
\end{eqnarray}

\noindent
where $m_0\equiv\langle\tilde{B}|H_0|\tilde{B}\rangle$ is the common core mass, while $F_8$ and $D_8$ ($F_3$ and $D_3$) represent the two reduced matrix elements for ${H_8}$ (${H_3}$). It is clear that our choice of ${H}$ guarantees the three sum rules

\begin{equation}
m_{\tilde{\Sigma}^{0}}=\frac{1}{2}
(m_{\tilde{\Sigma}^{+}}+m_{\tilde{\Sigma}^{-}}),
\label{11}
\end{equation}

\begin{equation}
m_{\tilde{p}}-m_{\tilde{n}}=(m_{\tilde{\Sigma}^{+}}-m_{\tilde{\Sigma}^{-}})-
(m_{\tilde{\Xi}^{0}}-m_{\tilde{\Xi}^{-}}),
\label{12}
\end{equation}
and

\begin{equation}
2(m_{\tilde{N}}+m_{\tilde{\Xi}})=3m_{\tilde{\Lambda}}+m_{\tilde{\Sigma}}
\label{13}
\end{equation}
where

\begin{equation}
m_{\tilde{N}}\equiv
\frac{1}{2}(m_{\tilde{p}}+m_{\tilde{n}}),\; m_{\tilde{\Xi}}\equiv
\frac{1}{2}(m_{\tilde{\Xi}^{0}}+m_{\tilde{\Xi}^{-}}),\;  m_{\tilde{\Sigma}}\equiv\frac{1}{3}(m_{\tilde{\Sigma}^{+}}+m_{\tilde{\Sigma}^{-}}+m_{\tilde{\Sigma}^0}), 
\label{13'}
\end{equation}
are the average masses of the isospin multiplets. Eqs.~(\ref{12}) and (\ref{13}) correspond to the Coleman-Glashow~\cite{10} and the Gell-Mann-Okubo~\cite{11} mass formulas for the core baryons. The physical baryon masses $m_B$ do not obey these two relations exactly due to the $SU(3)$ breaking in the wavefunction (Eq.~(\ref{1})) due to parameters $a(N)$, $b(N)$ for $N={\bf 1,10,\bar{10},27}$. However, Eq.~(\ref{11}) is obeyed by $m_{\Sigma^{\pm}}$ and $m_{\Sigma^0}$ since our wavefunction respects isospin.

As they stand, Eqs.~(\ref{10}) provide a model for the eight baryon masses $m_{\tilde B}$ in terms of five unknown $m_0$, $F_3$, $F_8$, $D_3$ and $D_8$. We can treat these five as independent parameters or try and connect them with the quark masses $m_q$ which enter in the baryon MM's through $\mu_q$. To do this, we note that the naive assumption that $m_{\tilde B}$ is equal to the sum of the masses of its three constituent quarks gives $m_{\tilde p}=2m_u+m_d$, $m_{\tilde\Lambda}=m_{\tilde{\Sigma}^0}=m_u+m_d+m_s$, $m_{\tilde{\Xi}^0}=2m_s+m_d$, etc. 
Such a model would amount to putting
  
\begin{eqnarray}
m_{0} &\equiv& m_{u}+m_{d}+m_{s},
\nonumber\\
F_{8} &\equiv& m_{s}-\frac{1}{2}(m_{u}+m_{d}),
\label{14}
\\
F_3 &\equiv&\frac{1}{2}(m_d-m_u),
\nonumber
\end{eqnarray}
with $D_3=D_8=0$ in Eqs.~(\ref{10}). Motivated by this observation, for our fits we also consider the alternative model for $m_{\tilde B}$ where

\begin{eqnarray}
m_{\tilde p}&=&2m_u+m_d-(D_8-D_3),
\nonumber\\
m_{\tilde n}&=&m_u+2m_d-(D_8+D_3),
\nonumber\\
m_{\tilde \Lambda}&=&m_u+m_d+m_s-2D_8,
\nonumber\\
m_{\tilde{\Sigma}^{+}}&=&2m_u+m_s+2D_8,
\nonumber\\
m_{\tilde{\Sigma}^{0}}&=&m_u+m_d+m_s+2D_8,
\label{15}
\\
m_{\tilde{\Sigma}^{-}}&=&2m_d+m_s+2D_8,
\nonumber\\
m_{\tilde \Xi^{0}}&=&m_u+2m_s-(D_8+D_3),
\nonumber\\
m_{\tilde \Xi^{-}}&=&m_d+2m_s-(D_8-D_3).
\nonumber
\end{eqnarray}
This model for $m_{\tilde B}$ treats $D_3$ and $D_8$ as extra independent parameters in fitting $m_B$ unlike the model for $m_{\tilde B}$ in Eqs.~(\ref{10}) which has five parameters. We will use both the models (Eqs.~(\ref{10}) and Eqs.~(\ref{15})) for $m_{\tilde B}$ to make simultaneous fits to baryon masses and MM's.

Since the actual baryon masses satisfy Eqs.~(\ref{11})--(\ref{13}) to a good accuracy, one may ask: why not fit the $m_B$ with five parameters as in Eq.~(\ref{10})?; and does the wavefunction or coefficients $\Omega_{B\tilde B}$ play a significant role?
The answers lie in the fact that a fit to the 8 physical masses $m_B$ using Eqs.~(\ref{10}) directly (with a theoretical error of 1 MeV) gives $\chi^2/DOF=50.37/3$.
Instead, the use of Eq.~(\ref{8}) for $m_B$ with $m_{\tilde B}$ given by Eqs.~(\ref{10}) gives very good fits to $m_{B}$ (see Secs.~\ref{iv} and \ref{v}).
So, the wavefunction parameters in $\Omega_{B\tilde B}$ do play a significant part.

\subsection{Semileptonic Decays (SLD's)}
\label{iiic}

The detailed expressions for $G_{V,A}(B\rightarrow B^\prime)=\langle B^\prime|J_{V,A}|B\rangle$ of the charge changing hadronic vector ($J_V$) and axial vector ($J_A$) currents using our wavefunction (Eq.~(\ref{2})) are given in Ref.~\cite{9}.
Here we briefly summarize how they were calculated.

The $\Delta S=0$ and $\Delta S=1$ vector currents are the total isospin raising ($I_{+}=I_{+}^{(q)}+I_{+}^{(s)}$) and $V$-spin lowering ($V_{-}=V_{-}^{(q)}+V_{-}^{(s)}$) operators~\cite{9}.
The operators $I_{+}^{(q)}$ and $V_{-}^{(q)}$ act on the quarks in the core baryons and $I_{+}^{(s)}$ and $V_{-}^{(s)}$ act on the sea states in the wavefunction.
However, the axial vector current has a quark part $J_{A}^{(q)}$ and a sea part $J_{A}^{(s)}$ which may, in general, have different relative strengths, so that

\begin{equation}
J_A(\Delta S=0,1)=J_{A}^{(q)}(\Delta S=0,1)+A_{0,1}J_{A}^{(s)}(\Delta S=0,1)
\label{16}
\end{equation}
where the constants $A_0$ and $A_1$ specify the strength of $J_{A}^{(s)}$ relative to $J_{A}^{(q)}$ for $\Delta S=0$ and $\Delta S=1$ transitions respectively.
In SQM, $J_{A}^{(q)}(\Delta S=0)=\sum_{q}I_{+}^{(q)}\sigma_{z}^{q}$ and $J_{A}^{(q)}(\Delta S=1)=\sum_{q}V_{-}^{(q)}\sigma_{z}^{q}$ so that, in analogy, we took $J_{A}^{(s)}(\Delta S=0)=2I_{+}^{(s)}S_{z}^{(s)}$ and $J_{A}^{(s)}(\Delta S=1)=2V_{-}^{(s)}S_{z}^{(s)}$ where $S_{z}^{(s)}$ is the spin operator acting only on the sea states in the wavefunction.
For $\Delta S=0$ transitions, the quark part was sufficient so that $A_0=0$ for all the fits.
For $\Delta S=1$ transitions, a direct sea contribution through $J_A^{(s)}$ is needed when the theoretical error on the MM's is very small. 

\section{Combined Fits and Results}
\label{iv}

In the last section, we have considered three possible models for $m_{\tilde B}$ (the core baryon masses) which could be used in Eq.~(\ref{8}). These are:

\paragraph*{\bf A)}

The naive or simple quark model assumption that the mass of baryon $\tilde B$ is equal to the sum of its three constituent quarks. Thus, all $m_{\tilde B}$ are given in terms of 3 $m_q$'s ($q=u,d,s$) and this corresponds to use of Eq.~(\ref{15}) with $D_3=D_8=0$. 
This model is attractive as it does not introduce new parameters for $m_{\tilde B}$ since the $m_q$'s enter as parameters in the MM's (see Sec.~\ref{iiia}).

\paragraph*{\bf B)}

The model for $m_{\tilde B}$ is given by Eq.~(\ref{15}) and introduces two new parameters $D_3$ and $D_8$ for eight $m_{\tilde B}$.
For the 3 average isospin multiplet masses (see Eq.~(\ref{13'})) $m_{\tilde N}$, $m_{\tilde \Sigma}$, $m_{\tilde\Xi}$ and $m_{\tilde\Lambda}$ this model has $D_8$ as an extra parameter. 

\paragraph*{\bf C)}

This model for $m_{\tilde B}$ is given by Eq.~(\ref{10}) and introduces 5 new parameters ($m_0$, $F_3$, $D_3$, $F_8$, $D_8$) for the $m_{\tilde B}$'s and three parameters ($m_0$, $F_8$, $D_8$) for the average isospin multiplet masses.

To test the viability of these models we made extensive and systematic fits to the 8 masses and 8 MM's using Eqs.~(\ref{5})--(\ref{8}).
For the fits we used theoretical errors added in quadratures to the experimental errors~\cite{12}.
For the MM's of the baryons we chose initially $0.1\mu_N$ and for the masses 1~MeV.
A motivation for adding these errors is that all masses and MM's are treated ``democratically"

\subsection{Fits for Masses and MM's}

We briefly summarize the main results \cite{13}.

\paragraph*{\bf Fit 1.}

Use of Model~A for the masses gives $\chi^2/DOF=146/9$ for 16 data for the best fit with seven parameters (4 for the sea and 3 $m_q$'s).

\paragraph*{\bf Fit 2.}

Use of Model~B for the masses improves the situation dramatically because of the parameters $D_3$ and $D_8$. This best fit gives $\chi^2/DOF=5.47/8$ with eight parameters (3 for the sea, 3 $m_q$'s, $D_3$, and $D_8$).

\paragraph*{\bf Fit 3.}

Use of Model~C gives $\chi^2/DOF=2.5/6$ with 10 parameters (2 for the sea, 3 $m_q$'s, and 5 ($m_0$, $F_3$, $F_8$, $D_3$, $D_8$) for the masses).

What we learn from the above is that Model~A is not viable. As we shall see, Model~B works only with a generous theoretical error like $0.1\mu_N$ for the MM's and later fits will use only Model~C which seems the most viable model for the core baryon masses.

\subsection{Results for Masses Using the Wavefunction Determined by Earlier Fits to MM's and SLD's}
 
In Ref.~\cite{9}, excellent 3 and 7 parameter fits were obtained to MM's and SLD's (12 data) using theoretical errors of 0.1$\,\mu_N$ and 0.001$\,\mu_N$ respectively. Since these fits specify the $m_q$'s and the wavefunction it is of interest to see their prediction for the masses. The prediction for the 4 average isospin multiplet masses using Eq.~(\ref{8}) and Model~C are given in Table~\ref{tabla4}.

As we can see, the prediction for the masses are fairly good. This encouraged us to make fresh combined fits to 8 MM's, 4 semileptonic decays (SLD's) and 8 masses using Model~C.

\subsection{Combined Fits to MM's, SLD's and Masses}

The combined fits here are different from those in Ref.~\cite{9} since, in addition to MM's and SLD's, we also fit the baryon masses. For the masses and MM's Eqs.~(\ref{5})--(\ref{8}) were used together with Model~C for the core baryon masses. The $G_V$ and $G_A$ for the SLD's depend on the sea parameters and they were calculated as described briefly in Sec.~\ref{iiic}. For explicit expressions for $G_V$ and $G_A$ see Ref.~\cite{9}.

For $\Delta S=0$ transitions, the quark part was sufficient so that $A_0=0$ for all the fits. For $\Delta S=1$ transitions, a direct sea contribution through $J_{A}^{(s)}$ is needed when the theoretical error on the MM's is very small. In Table~\ref{tabla5}, $A_1=0$ for theoretical error of 0.1$\,\mu_N$ (fit in Column~3) while for theoretical errors of 0.01$\,\mu_N$ and 0.001$\,\mu_N$ (fits in Columns~4 and 5) we took $A_1=-1$. These values for $A_0$ and $A_1$ can be treated as input values since varying them does not affect the fits too much.

Fits to 20 pieces of data were made using theoretical errors of 0.1$\,\mu_N$, 0.01$\,\mu_N$ and 0.001$\,\mu_N$ for the MM's. In each case, a theoretical error of 1 MeV was used for the masses, while experimental errors were used for $G_A/G_V$ for the SLD's. The results are displayed in Table~\ref{tabla5} and the values of the parameters are compared in Table~\ref{tabla6}.

Of the many good fits possible, Table~\ref{tabla5} displays fits which have a reasonable $\chi^2/\hbox{DOF}$ with as few parameters as possible\footnote{Note, very much lower $\chi^2$ can be achieved for fits in Columns~4 and 5 of Table~\ref{tabla5} with more parameters for the sea.}. The number of parameters describing the sea increase from 2 for a large theoretical error of 0.1$\,\mu_N$ to 6 for small error of 0.001$\,\mu_N$. Our fit for 0.1$\,\mu_N$ theoretical error is comparable to other phenomenological fits which use this error to fit MM's and SLD's alone. Fits for the extremely small theoretical error, {\em e.g.}
0.001$\,\mu_N$ (close to most experimental errors) are not given by other models. In contrast, our wavefunction gives a very good fit (Column~5 of Table~\ref{tabla5}) suggesting that our phenomenological model for incorporation of the sea may be in the right direction.

Most of the $\chi^2$ in the 0.1$\,\mu_N$ fit is from SLD's. Actual break up for 0.1$\,\mu_N$ is $\chi^2_{\rm MM}=2.61$, $\chi^2_{\rm Masses}=0.75$, $\chi^2_{\rm SLD}=6.90$, while for 0.001$\,\mu_N$ it is $\chi^2_{\rm MM}=0.81$, $\chi^2_{\rm Masses}=0.62$, $\chi^2_{\rm SLD}=1.23$.

Comparison of the parameters for 3 fits in Table~\ref{tabla6} shows the following:

a) The values of the quark masses which enter in the MM's are approximately the same. For smaller theoretical error on MM's the data requires $m_u>m_d$. The $m_q$'s obtained do not satisfy Eq.~(\ref{14}) ruling out Model B for the core baryon masses. 

b) The 5 parameters of Model C for core baryon masses are approximately the same. It is interesting that the average core baryon octet masses $m_0\approx 1159$ MeV is close to the experimental value $\frac{1}{8}\sum_{B}m_B\approx 1151$ MeV. 

c) Since the $m_q$ needed for the MM's do not satisfy Eq.~(\ref{14}), it is clear that the sea is responsible (through Eq.~(\ref{8})) for a good fit to the physical baryon masses.

d) One requires more parameters to describe the sea as one reduces the theoretical error on the MM's. The extra parameters are connected with the vector sea component in the wavefunction.

e) $SU(3)$ flavor breaking effects in all the fits are mainly given by the scalar sea parameter $a({\bf 10})$ which contributes only to the $\Sigma^{\pm,0}$ and $\Xi^0$, $\Xi^-$ MM's and masses. For smaller theoretical error, breaking effects through the vector sea parameter $b({\bf 1})$ (which contributes only to $\mu_\Lambda$ and $m_\Lambda$) are needed.

Table~\ref{tabla6} also gives the values of $(\Delta q)^p$ which are relevant for the measured [1,16] nucleon spin distribution discussed below. The interesting point to note is that in all fits $(\Delta u)^p\approx 1$, $(\Delta d)^p\approx$ $-$0.26 to $-$0.3 with quite small $(\Delta s)^p\approx 0.01$. This in contrast to other models [14,15] which require $(\Delta u)^p\approx 0.8$, $(\Delta d)^p\approx -0.5$, $(\Delta s)^p\approx -0.15$. Thus, our fits need only a small strange-quark content in the nucleon and thus, are physically quite different to the other phenomenological fits in the literature.

\section{Spin Distributions}
\label{v}

The spin distribution, $I_{1B}$, for baryon $B$ is defined as

\begin{equation}
I_{1B}\equiv\int_{0}^{1}g_{1B}(x)dx,
\label{18}
\end{equation}
where the spin structure function $g_{1B}$ occurs in polarized electron-baryon scattering.

In SQM, $I_{1B}$ is given by the expectation value $I_{1B}\equiv\langle B|\hat{I}_{1}^{(q)}|B\rangle$ where the quark operator $\hat{I}_{1}^{(q)}=(1/2)\sum_{q}e_{q}^2\sigma_{Z}^{q}$.
This gives

\begin{equation}
I_{1B}^{(q)}=\frac{1}{18}\left [ 4(\Delta u)^B+(\Delta d)^B+(\Delta s)^B\right ].
\label{19}
\end{equation}
In our model in addition to the quarks there can be a direct sea contribution $I_{1B}\equiv\langle B|\hat{I}_{1B}^{(s)}|B\rangle$ where by analogy we take $\hat{I}_{1B}^{(s)}=e_s^2S_{Z}^{(s)}$. Thus only the charged states in the vector sea will contribute to $I_{1B}^{(s)}$.
For the nucleons, one obtains

\begin{equation}
I_{1p}^{(s)}=\frac{1}{3N_1^2}\left (\bar{\beta}_{2}^{\prime 2}+\frac{2}{3} \bar{\beta}_{3}^{\prime 2}+\frac{1}{3}\bar{\beta}_{4}^{\prime 2}\right ),\qquad
I_{1n}^{(s)}=\frac{1}{3N_1^2}\left (\frac{2}{3} \bar{\beta}_{3}^{\prime 2}+\frac{2}{3}\bar{\beta}_{4}^{\prime 2}\right ).
\label{20}
\end{equation}
Putting the two contributions together we have

\begin{equation}
I_{1B}=I_{1B}^{(q)}+B_{1}I_{1B}^{(s)},
\label{21}
\end{equation}
where $B_1$ determines the strength of the direct sea contribution to the valence quark contribution. Since the value of $B_1$ is not known \'a priori, so phenomenologically it may be treated as a parameter.

Experiment [1,16] gives $I_{1p}=0.126\pm 0.018$ and $I_{1n}=-0.08\pm 0.06$ which are very different from the SQM predictions $I_{1p}=5/18=0.2778$ and $I_{1n}=0$. One must note that the EMC experiment gives $I_{1p}$ for $\langle Q^2\rangle =10.7\, (\rm{GeV}/\rm{c})^2$ and this could be very different for the very low $Q^2$ ($\approx 0$) result predicted by SQM or other theoretical models. This could mean that a model which gives values for $I_{1B}$ differing by 2--3 standard deviations from experiment may be quite acceptable. 

Using the values for $(\Delta q)^p$ in Table~\ref{tabla6} it is clear that our values for $I_{1p}^{(q)}$ ($\approx 0.2$) are much lower than the SQM value but still $4\sigma$ higher than experiment. This may be due to the large $\langle Q^2\rangle$ in the experiment. In our model, in addition to the quark part $I_{1B}^{(q)}$ one can invoke the direct sea contribution $I_{1B}^{(s)}$. The numerical values are listed in Table~\ref{tabla7} with the choice $B_{1}=-1$. As one can see, one obtains good agreement with experiment only for the fit (second column, Table~\ref{tabla7}) when extremely large theoretical error for the magnetic moments was used. 

\section{Summary}
\label{vi}

We have shown that our wavefunction, for spin 1/2 baryons, which incorporates a flavor octet sea component can simultaneously give a good fit to their magnetic moments, weak decays constants $G_A/G_V$ for both $\Delta S=0,1$ semileptonic decays as well as the eight baryon masses. In addition, these fits give viable predictions for the nucleon spin distributions. The sea was found to be both scalar (spin 0) and vector (spin 1). The $SU(3)$ flavor breaking in the wavefunction is mainly due to the scalar component. Two important features of the fits are that the valence quarks carry about 70\% of the proton spin and that the nucleons have a small strange-quark content. 

In conclusion, our model can account for all the static properties of the eight low-lying spin 1/2 baryons.

\acknowledgments

This work was partially supported by CONACyT (M\'exico).

\begin{table}
\caption{
Contribution to the physical baryon state $B(Y,I,I_3)$ formed out of
$\tilde{B}(Y,I,I_3)$ and flavor octet states $S(Y,I,I_3)$ (see third term in Eq.~(1)).
The core baryon states $\tilde{B}$ denoted by $\tilde{p}$, $\tilde{n}$, etc.
are the normal 3 valence quark states of SQM.
The sea octet states are denoted by $S_{\pi^+}=S(0,1,1)$, etc. as in Eq.~(3).
Further, $(\tilde{N}S_{\pi})_{I,I_3}$, $(\tilde{\Sigma}S_{\bar{K}})_{I,I_3}$,
$(\tilde{\Sigma}S_{\pi})_{I,I_3}$, $\dots$ stand for total $I$, $I_3$
{\em normalized} combinations of $\tilde{N}$ and $S_{\pi}$, etc. Only the contribution from the third term is given. Fourth term has exactly the same flavor symmetries with coefficients $(\bar{\beta}_i, \beta_i, \gamma_i, \delta_i)\rightarrow(\bar{\beta}_{i}^\prime, \beta_{i}^\prime, \gamma_{i}^\prime, \delta_{i}^\prime)$. See Table~\ref{tabla2} for the coefficients $\bar{\beta}_i$, $\beta_i$, $\gamma_i$,
and $\delta_i$.
}~
\label{tabla1}
\begin{tabular}
{
cc
}
$B(Y,I,I_3)$ &
$\tilde{B}(Y,I,I_3)$ and $S(Y,I,I_3)$
\\
\hline
\\
$p$ &
$
\bar{\beta}_1 \tilde{p} S_{\eta}
+ \bar{\beta}_2 \tilde{\Lambda} S_{K^+}
+ \bar{\beta}_3 (\tilde{N}S_{\pi})_{1/2,1/2}
+ \bar{\beta}_4 (\tilde{\Sigma}S_{K})_{1/2,1/2}
$
\\
\\
$n$ &
$
\bar{\beta}_1 \tilde{n} S_{\eta}
+ \bar{\beta}_2 \tilde{\Lambda} S_{K^0}
+ \bar{\beta}_3 (\tilde{N}S_{\pi})_{1/2,-1/2}
+ \bar{\beta}_4 (\tilde{\Sigma}S_{K})_{1/2,-1/2}
$
\\
\\
$\Xi^0$ &
$
\beta_1 \tilde{\Xi}^0 S_{\eta}
+ \beta_2 \tilde{\Lambda} S_{\bar{K}^0}
+ \beta_3 (\tilde{\Xi}S_{\pi})_{1/2,1/2}
+ \beta_4 (\tilde{\Sigma}S_{\bar{K}})_{1/2,1/2}
$
\\
\\
$\Xi^-$ &
$
\beta_1 \tilde{\Xi}^- S_{\eta}
+ \beta_2 \tilde{\Lambda} S_{\bar{K}^-}
+ \beta_3 (\tilde{\Xi}S_{\pi})_{1/2,-1/2}
+ \beta_4 (\tilde{\Sigma}S_{\bar{K}})_{1/2,-1/2}
$
\\
\\
$\Sigma^+$ &
$
\gamma_1 \tilde{p} S_{\bar{K}^0}
+ \gamma_2 \tilde{\Xi}^0 S_{K^+}
+ \gamma_3 \tilde{\Lambda} S_{\pi^+}
+ \gamma_4 \tilde{\Sigma}^+ S_{\eta}
+ \gamma_5 (\tilde{\Sigma}S_{\pi})_{1,1}
$
\\
\\
$\Sigma^-$ &
$
\gamma_1 \tilde{n} S_{K^-}
+ \gamma_2 \tilde{\Xi}^- S_{K^0}
+ \gamma_3 \tilde{\Lambda} S_{\pi^-}
+ \gamma_4 \tilde{\Sigma}^- S_{\eta}
+ \gamma_5 (\tilde{\Sigma}S_{\pi})_{1,-1}
$
\\
\\
$\Sigma^0$ &
$
\gamma_1 (\tilde{N} S_{\bar{K}})_{1,0}
+ \gamma_2 (\tilde{\Xi} S_{K})_{1,0}
+ \gamma_3 \tilde{\Lambda} S_{\pi^0}
+ \gamma_4 \tilde{\Sigma}^0 S_{\eta}
+ \gamma_5 (\tilde{\Sigma}S_{\pi})_{1,0}
$
\\
\\
$\Lambda$ &
$
\delta_1 (\tilde{N} S_{\bar{K}})_{0,0}
+ \delta_2 (\tilde{\Xi} S_{K})_{0,0}
+ \delta_3 \tilde{\Lambda} S_{\eta}
+ \delta_4 (\tilde{\Sigma}S_{\pi})_{0,0}
$
\\
\end{tabular}
\end{table}

\begin{table}
\squeezetable
\caption{
The coefficients $\bar{\beta}_i$, $\beta_i$, $\gamma_i$, and $\delta_i$ in
Table~\ref{tabla1} expressed in terms of the coefficients $a(N)$,
$N={\bf 1, 8_{F}, 8_{D}, 10, \bar{10}, 27}$, in the $3^{\rm rd}$ term
(from scalar sea) in Eq.~(1).
The corresponding coefficients $\bar{\beta'}_i$, $\beta'_i$, $\gamma'_i$,
and $\delta'_i$ determining the flavor structure of $4^{\rm th}$ term in
Eq.~(1) can be obtained from $\bar{\beta}_i$, etc. by the replacement
$a(N)\rightarrow b(N)$ (see text).
}~
\label{tabla2}
\begin{tabular}{cc}
\\
$
\bar{\beta}_1=\frac{1}{\sqrt {20}}(3a({\bf 27})-a({\bf 8_D}))+
\frac{1}{2}(a({\bf 8_F})+a({\bf \bar{10}}))
$
&
$
\beta_1=\frac{1}{\sqrt {20}}(3a({\bf 27})-a({\bf 8_D}))-
\frac{1}{2}(a({\bf 8_F})-a({\bf 10}))
$
\\
\\
$
\bar{\beta}_2=\frac{1}{\sqrt {20}}(3a({\bf 27})-a({\bf 8_D}))-
\frac{1}{2}(a({\bf 8_F})+a({\bf \bar{10}}))
$
&
$
\beta_2=\frac{1}{\sqrt {20}}(3a({\bf 27})-a({\bf 8_D}))+
\frac{1}{2}(a({\bf 8_F})-a({\bf 10}))
$
\\
\\
$
\bar{\beta}_3=\frac{1}{\sqrt {20}}(a({\bf 27})+3a({\bf 8_D}))+
\frac{1}{2}(a({\bf 8_F})-a({\bf \bar{10}}))
$
&
$
\beta_3=-\frac{1}{\sqrt {20}}(a({\bf 27})+3a({\bf 8_D}))+
\frac{1}{2}(a({\bf 8_F})+a({\bf 10}))
$
\\
\\
$
\bar{\beta}_4 =-\frac{1}{\sqrt {20}}(a({\bf 27})+3a({\bf 8_D}))+
\frac{1}{2}(a({\bf 8_F})-a({\bf \bar{10}}))
$
&
$
\beta_4=\frac{1}{\sqrt{20}}(a({\bf 27})+3a({\bf 8_D}))+
\frac{1}{2}(a({\bf 8_F})+a({\bf 10}))
$
\\
\\
$
\gamma_1=\frac{1}{\sqrt{10}}(\sqrt{2}a({\bf 27})-\sqrt{3}a({\bf 8_D}))
$
&
$
\delta_1=\frac{1}{\sqrt{20}}(\sqrt{3}a({\bf 27})+\sqrt{2}a({\bf 8_D}))
$
\\
\\
$
+\frac{1}{\sqrt{6}}(a({\bf 8_F})-a({\bf 10})+a({\bf\bar{10}}))
$
&
$
+\frac{1}{2}(\sqrt{2}a({\bf 8_F})+a({\bf 1}))
$
\\
\\
$
\gamma_2=\frac{1}{\sqrt{10}}(\sqrt{2}a({\bf 27})-\sqrt{3}a({\bf 8_D}))
$
&
$
\delta_2=-\frac{1}{\sqrt{20}}(\sqrt{3}a({\bf 27})+\sqrt{2}a({\bf 8_D}))
$
\\
\\
$
-\frac{1}{\sqrt{6}}(a({\bf 8_F})-a({\bf 10})+a({\bf\bar{10}}))
$
&
$
+\frac{1}{2}(\sqrt{2}a({\bf 8_F})-a({\bf 1}))
$
\\
\\
$
\gamma_3=\frac{1}{\sqrt{10}}(\sqrt{3}a({\bf 27})+\sqrt{2}a({\bf 8_D}))-
\frac{1}{2}(a({\bf 10})+a({\bf\bar{10}}))
$
&
$
\delta_3=\frac{3\sqrt{3}}{\sqrt{40}}a({\bf 27})-
\frac{1}{\sqrt{5}}a({\bf 8_D})-
\frac{\sqrt{2}}{4}a({\bf 1})
$
\\
\\
$
\gamma_4=\frac{1}{\sqrt{10}}(\sqrt{3}a({\bf 27})+\sqrt{2}a({\bf 8_D}))+
\frac{1}{2}(a({\bf 10})+a({\bf\bar{10}}))
$
&
$
\delta_4=-\frac{1}{\sqrt{40}}a({\bf 27})-\sqrt{\frac{3}{5}}a({\bf 8_D})+
\frac{\sqrt{6}}{4}a({\bf 1})
$
\\
\\
$
\gamma_5=\frac{1}{\sqrt{6}}(2a({\bf 8_F})+a({\bf 10})-a({\bf\bar{10}}))
$
&
\\
\end{tabular}
\end{table}

\begin{table}
\caption{
The coefficients $\Omega_{B\tilde B}$ in Eq.~(\ref{8}).
Here $\Omega_{B\tilde B}^\prime=N_i^2 \Omega_{B\tilde B}$, where $N_i$ ($i=1,2,3,4$) are the normalization constants for the $(p,n)$, $(\Xi^0,\Xi^{-})$, $(\Sigma^{\pm},\Sigma^0)$, and $\Lambda^0$ isospin multiplets, respectively:
$N^2_1=N^2_0+a^2({\bf\bar{10}})+b^2({\bf\bar{10}})$,
$N^2_2=N^2_0+a^2({\bf 10})+ b^2({\bf10})$,
$N^2_3=N^2_0+\sum_{N={\bf 10,\bar{10}}}[a^2(N)+b^2(N)]$,
and
$N^2_4=N^2_0+a^2({\bf 1})+b^2({\bf 1})$,
with
$N^2_0=1+b^2_0+\sum_{N={\bf 8_D,8_F,27}}[a^2(N)+b^2(N)]$.
}~
\label{tabla3}
\begin{tabular}{ll}
\\
$\Omega^\prime_{p\tilde{p}}=\Omega^\prime_{n\tilde{n}}=
1+b_0^2+\bar{\beta}_{1}^{2}+\frac{1}{3}\bar{\beta}_{3}^{2}
+\bar{\beta}_{1}^{\prime 2}+\frac{1}{3}\bar{\beta}_{3}^{\prime 2}$
&
$\Omega^\prime_{p\tilde{n}}=\Omega^\prime_{n\tilde{p}}=
\frac{2}{3}(\bar{\beta}_{3}^{2}+\bar{\beta}_{3}^{\prime 2})$ 
\\
&
$\Omega^\prime_{p\tilde{\Lambda}}=\Omega^\prime_{n\tilde{\Lambda}}=
\bar{\beta}_{2}^{2}+\bar{\beta}_{2}^{\prime 2}$
\\
$\Omega^\prime_{p\tilde{\Sigma}^{+}}=\Omega^\prime_{n\tilde{\Sigma}^{-}}=
\frac{2}{3}(\bar{\beta}_{4}^{2}+\bar{\beta}_{4}^{\prime 2})$
&
$\Omega^\prime_{p\tilde{\Sigma}^{0}}=\Omega^\prime_{n\tilde{\Sigma}^{0}}=
\frac{1}{3}(\bar{\beta}_{4}^{2}+\bar{\beta}_{4}^{\prime 2})$
\\ 
$\Omega^\prime_{p\tilde{\Sigma}^{-}}=\Omega^\prime_{n\tilde{\Sigma}^{+}}=0$
&  
$\Omega^\prime_{p\tilde{\Xi}^{0}}=\Omega^\prime_{n\tilde{\Xi}^{0}}=
\Omega^\prime_{p\tilde{\Xi}^{-}}=\Omega^\prime_{n\tilde{\Xi}^{-}}=0$ 
\\
&
\\
$\Omega^\prime_{\Xi^{0}\tilde{p}}=\Omega^\prime_{\Xi^{-}\tilde{n}}=
\Omega^\prime_{\Xi^{0}\tilde{n}}=\Omega^\prime_{\Xi^{-}\tilde{p}}=0$
& $\Omega^\prime_{\Xi^{0}\tilde{\Lambda}}=\Omega^\prime_{\Xi^{-}\tilde{\Lambda}}=
\beta_{2}^{2}+\beta_{2}^{\prime 2}$
\\
$\Omega^\prime_{\Xi^{0}\tilde{\Sigma}^{+}}=
\Omega^\prime_{\Xi^{-}\tilde{\Sigma}^{-}}=
\frac{2}{3}(\beta_{4}^{2}+\beta_{4}^{\prime 2})$
&
$\Omega^\prime_{\Xi^{0}\tilde{\Sigma}^{0}}=
\Omega^\prime_{\Xi^{-}\tilde{\Sigma}^{0}}=
\frac{1}{3}(\beta_{4}^{2}+\beta_{4}^{\prime 2})$
\\ 
$\Omega^\prime_{\Xi^{0}\tilde{\Sigma}^{-}}=
\Omega^\prime_{\Xi^{-}\tilde{\Sigma}^{+}}=0$
& 
$\Omega^\prime_{\Xi^{0}\tilde{\Xi}^{0}}=\Omega^\prime_{\Xi^{-}\tilde{\Xi}^{-}}=
1+b_0^2+\beta_{1}^{2}+\frac{1}{3}\beta_{3}^{2}
+\beta_{1}^{\prime 2}+\frac{1}{3}\beta_{3}^{\prime 2}$
\\
$\Omega^\prime_{\Xi^{0}\tilde{\Xi}^{-}}=\Omega^\prime_{\Xi^{-}\tilde{\Xi}^{0}}=
\frac{2}{3}(\beta_{3}^{2}+\beta_{3}^{\prime 2})$
&
\\
&
\\
$\Omega^\prime_{\Sigma^{+}\tilde{p}}=\Omega^\prime_{\Sigma^{-}\tilde{n}}=
\gamma_{1}^{2}+\gamma_{1}^{\prime 2}$
&
$\Omega^\prime_{\Sigma^{+}\tilde{n}}=\Omega^\prime_{\Sigma^{-}\tilde{p}}=0$
\\
$\Omega^\prime_{\Sigma^{+}\tilde{\Lambda}}=
\Omega^\prime_{\Sigma^{-}\tilde{\Lambda}}=
\gamma_{3}^{2}+\gamma_{3}^{\prime 2}$
&
$\Omega^\prime_{\Sigma^{+}\tilde{\Sigma}^{+}}=
\Omega^\prime_{\Sigma^{-}\tilde{\Sigma}^{-}}=
1+b_0^2+\gamma_{4}^{2}+\frac{1}{2}\gamma_{5}^{2}
+\gamma_{4}^{\prime 2}+\frac{1}{2}\gamma_{5}^{\prime 2}$
\\
$\Omega^\prime_{\Sigma^{+}\tilde{\Sigma}^{0}}=
\Omega^\prime_{\Sigma^{-}\tilde{\Sigma}^{0}}=
\frac{1}{2}(\gamma_{5}^{2}+\gamma_{5}^{\prime 2})$
&
\\
$\Omega^\prime_{\Sigma^{+}\tilde{\Xi}^{0}}=
\Omega^\prime_{\Sigma^{-}\tilde{\Xi}^{-}}=
\gamma_{2}^{2}+\gamma_{2}^{\prime2}$
&
$\Omega^\prime_{\Sigma^{+}\tilde{\Sigma}^{-}}=
\Omega^\prime_{\Sigma^{-}\tilde{\Sigma}^{-}}=0$
\\
$\Omega^\prime_{\Sigma^{+}\tilde{\Xi}^{-}}=
\Omega^\prime_{\Sigma^{+}\tilde{\Xi}^{-}}=0$
&
$\Omega^\prime_{\Sigma^{0}\tilde{B}}=
\frac{1}{2}(\Omega^\prime_{\Sigma^{+}\tilde{B}}
+\Omega^\prime_{\Sigma^{-}\tilde{B}})$
\\
&
\\
$\Omega^\prime_{\Lambda\tilde{p}}=\Omega^\prime_{\Lambda\tilde{n}}=
\frac{1}{2}(\delta_{1}^{2}+\delta_{1}^{\prime 2})$
&
$\Omega^\prime_{\Lambda\tilde{\Lambda}}=
1+b_0^2+\delta_{3}^{2}+\delta_{3}^{\prime 2}$
\\
$\Omega^\prime_{\Lambda\tilde{\Sigma}^{+}}=
\Omega^\prime_{\Lambda\tilde{\Sigma}^{0}}=
\Omega^\prime_{\Lambda\tilde{\Sigma}^{-}}=
\frac{1}{3}(\delta_{4}^{2}+\delta_{4}^{\prime 2})$
&
$\Omega^\prime_{\Lambda\tilde{\Xi}^{0}}=
\Omega^\prime_{\Lambda\tilde{\Xi}^{-}}=
\frac{1}{2}(\delta_{2}^{2}+\delta_{2}^{\prime 2})$
\\
\end{tabular}
\end{table}

\begin{table}
\caption{
Results of the fits for the average masses of the baryon isospin multiplets, $m_N$, $m_\Xi$, $m_\Sigma$, and $m_\Lambda$ with the sea and quark mass parameters given in Ref.~[9].
The masses were fitted adding a theoretical error of $2$MeV in quadratures to experimental errors.
We also give the values of the mass parameters $m_0$, $F_8$, and $D_8$.
}~
\label{tabla4}
\begin{tabular}{lr@{$\,\pm\,$}lrr}
Average isospin&\multicolumn{2}{c}{Data}&\multicolumn{2}{c}{Prediction}\\
\cline{4-5}
multiplet mass&\multicolumn{2}{c}{(MeV)}&
$0.1\mu_N$ &
$0.001\mu_N$
\\
\hline
$m_N$&
938.91897 & 0.00028 &
939.558 &
940.636

\\
$m_{\Lambda}$&
1115.684 & 0.006 &
1114.73 &
1113.19 

\\
$m_{\Sigma}$ &
1193.1118 & 0.111 &
1192.74 &
1192.07 
\\
$m_{\Xi}$&
1318.11 & 0.61 &
1318.87 &
1320.01

\\
\\
$m_0$ &
\multicolumn{2}{c}{}&
1157.46 &
1157.13

\\
$F_8$ &
\multicolumn{2}{c}{} &
223.65 &
213.72

\\
$D_8$ & 
\multicolumn{2}{c}{} &
31.20 &
23.21

\\
\\
$\chi^2({\rm masses})$ &
\multicolumn{2}{c}{} &
0.50 &
3.48

\\
$\chi^2(\hbox{MM's+SLD's})$ &
\multicolumn{2}{c}{} &
10.70 &
1.02

\\
$\chi^2{\rm (total)/DOF}$ &
\multicolumn{2}{c}{} &
11.20/10 &
4.50/6

\\
\end{tabular}
\end{table}

\begin{table}
\caption{
Results of the combined fits for MM's (in $\mu_N$), Masses (in MeV), and SLD's. The MM's were fitted with three different theoretical errors: $0.1\mu_N$, $0.01\mu_N$, and $0.001\mu_N$.
In each case, the masses were fitted with a theoretical error of 1~MeV, while experimental errors were used for the SLD's.
The values of the fitted parameters are given in Table~\ref{tabla6}.
For the fit in Column~3, input values $m_u=m_d=\frac{2}{3}m_s$ were used.
Note, fits with $m_u=m_d$ for theoretical errors of $0.01\mu_N$ and $0.001\mu_N$ are possible and give $\chi^2/\hbox{DOF}$ of 9.78/8 and 7.01/7 respectively.
}~
\label{tabla5}
\begin{tabular}{lr@{\,$\pm$\,}lrrr}
Quantity &
\multicolumn{2}{c}{Data}&
$0.1\mu_N$&
$0.01\mu_N$&
$0.001\mu_N$
\\
\hline

$\mu_p$ &
2.79284739 & $6\times 10^{-8}$ &
2.78188 &
2.79611 &
2.79286

\\
$\mu_n$ &
$-$1.9130428 & $5\times 10^{-7}$ &
$-$1.95706 &
$-$1.91252 &
$-$1.91306

\\
$\mu_{\Lambda}$ &
$-$0.613 & 0.004 &
$-$0.668 &
$-$0.623 &
$-$0.613

\\
$\mu_{\Sigma^+}$ &
2.458 & 0.010 &
2.522 &
2.453 &
2.457

\\
$\mu_{\Sigma^0}$ &
\multicolumn{2}{c}{--------} &
0.706 &
0.649 &
0.656

\\
$\mu_{\Sigma^-}$ &
$-$1.160 & 0.025 &
$-$1.110 &
$-$1.155 &
$-$1.146

\\
$\mu_{\Xi^0}$ &
$-$1.250 & 0.014 &
$-$1.157 &
$-$1.244 &
$-$1.250

\\
$\mu_{\Xi^-}$ &
$-$0.650 & 0.0025 &
$-$0.650 &
$-$0.646 &
$-$0.651

\\
$|\mu_{\Sigma^0\Lambda}|$ &
1.61 & 0.08 &
1.51 &
1.55 &
1.55

\\
\\
$m_p$&
938.27231 & 0.00028 &
938.362 &
938.283 &
938.126

\\
$m_n$&
939.56563 & 0.00028 &
939.747 &
939.705 &
939.528

\\
$m_{\Lambda}$ &
1115.684 & 0.006 &
1115.28 &
1115.44 &
1115.95

\\
$m_{\Sigma^+}$ &
1189.37 & 0.07 &
1189.08 &
1189.13 &
1189.18

\\
$m_{\Sigma^0}$ &
1192.55 & 0.08 &
1193.06 &
1193.09 &
1193.16

\\
$m_{\Sigma^-}$ &
1197.436 & 0.033 &
1197.05 &
1197.06 &
1197.14

\\
$m_{\Xi^0}$ &
1314.9 & 0.6 &
1315.03 &
1314.92 &
1314.69

\\
$m_{\Xi^-}$ &
1321.32 & 0.13 &
1321.52 &
1321.48 &
1321.28

\\
\\
$G_A/G_V(n\rightarrow p)$ &
1.2601 & 0.0025 &
1.2598 &
1.2599 &
1.2602

\\
$G_A/G_V(\Lambda \rightarrow p)$ &
0.718 & 0.015 &
0.739 &
0.726 &
0.719

\\
$G_A/G_V(\Sigma^{-}\rightarrow n)$ &
$-0.340$ & 0.017 &
$-0.304$ &
$-0.338$ &
$-0.339$

\\
$G_A/G_V(\Xi^{-}\rightarrow \Lambda)$ &
0.25 & 0.05 &
0.22 &
0.20 &
0.19

\\
\\
Inputs: $\quad\; A_0$ & \multicolumn{2}{c}{} & 0 & 0 & 0

\\
$\;\;\,\qquad\qquad A_1$ & \multicolumn{2}{c}{} & 0 & $-1$ & $-1$

\\
\\
$\chi^2/\hbox{DOF}$ &
\multicolumn{2}{c}{--------} &
10.27/12 &
3.91/7 &
2.66/6

\\
\end{tabular}
\end{table}

\begin{table}
\caption{
Comparison of the sea and mass parameters for the three fits shown in Table~\ref{tabla5} and the values of $(\Delta q)^p$.
The parameters $a({\bf 8_F})$ and $a({\bf 10})$ refer to a scalar sea while $b_0$, $b({\bf 1})$, $b({\bf 8_D})$ and $b({\bf 8_F})$, and $b({\bf\bar{10}})$ refer to a vector sea.
Note, $m_0$, $F_3$, $F_8$, $D_3$, $D_8$ parametrize the core baryon masses (See Eq.~(\ref{10})).
See text for details and Table~\ref{tabla5} for results.
}~
\label{tabla6}
\begin{tabular}{lrrr}
Parameter&0.1$\,\mu_N$&0.01$\,\mu_N$&0.001$\,\mu_N$
\\
\hline

$m_u$ &
249.435 &
253.445 &
253.346

\\
$m_d$ &
249.435 &
246.429 &
247.218

\\
$m_s$ &
374.153 &
382.796 &
379.122

\\
\\
$m_0$&
1158.73 &
1159.42 &
1159.23
\\
$F_8$&
225.263 &
208.397 &
208.723

\\
$D_8$&
31.728 &
26.688 &
26.797

\\
$F_3$&
2.491 &
2.275 &
2.286

\\
$D_3$ &
2.041 &
1.670 &
1.691
\\
\\
$a({\bf 8_F})$ &
----- & 
$-$0.1378 &
$-$0.1297

\\
$a({\bf 10})$ &
0.5922 & 
0.5183 &
0.5201

\\
$b_0$ &
----- &
0.3869 &
0.3816

\\
$b({\bf 1})$ &
----- &
----- &
0.0960

\\
$b({\bf 8_D})$ &
0.5658 &
0.2438 &
0.2827

\\
$b({\bf 8_F})$ &
----- &
$-$0.1443 &
$-$0.1050

\\
\\
$(\Delta u)^p$ &
0.9642 &
1.0035 &
1.0006

\\
$(\Delta d)^p$ &
$-$0.2956 &
$-$0.2564 &
$-$0.2597

\\
$(\Delta s)^p$ &
0.0081 &
0.0074 &
0.0074

\\
\end{tabular}
\end{table}

\begin{table}
\caption{
Predictions for nucleon spin distributions $I_{1p}$ and $I_{1n}$ (for 3 cases, see Table~\ref{tabla5} and Table~\ref{tabla6}) using the combined fits given in Sec.~\ref{iv}. The predictions are based on Eq.~(\ref{21}) for the choice $B_{1}=-1$.
}~
\label{tabla7}
\begin{tabular}{ccrrr}
Data&&0.1$\,\mu_N$&0.01$\,\mu_N$&0.001$\,\mu_N$\\
\hline

&$I_{1p}^{(q)}$&
0.198 &
0.209 &
0.208

\\
$I_{1p}=0.126\pm 0.018$&
$I_{1p}^{(s)}$ &
0.080 &
0.013 &
0.017

\\
&$I_{1p}$ &
0.118&
0.196 &
0.191

\\
\\
&$I_{1n}^{(q)}$&
$-$0.01 &
$-$0.001 &
$-$0.001

\\
$I_{1n}=-0.08\pm 0.06$&
$I_{1n}^{(s)}$ &
0.09 &
0.02 &
0.02

\\
&$I_{1n}$ &
$-$0.10 &
$-$0.02 &
$-$0.02

\\
\end{tabular}
\end{table}

\end{document}